\documentclass[aps,prl,twocolumn,superscriptaddress,10pt]{revtex4-2}
\usepackage[english]{babel}
\usepackage[utf8]{inputenc}
\usepackage{amsthm,amssymb}
\usepackage{mathtools}
\usepackage{physics}
\usepackage{xcolor}
\usepackage{graphicx}
\usepackage{siunitx}
\usepackage[left=23mm, right=13mm, top=35mm, columnsep=15pt]{geometry}
\usepackage{adjustbox}
\usepackage{placeins}
\usepackage[T1]{fontenc}
\usepackage{lipsum}
\usepackage{csquotes}
\usepackage[compat=1.0.0]{tikz-feynman}
\usepackage{feynmp-auto}
\usepackage[version=4]{mhchem}
\usepackage{prettyref}
\usepackage[pdftex, pdftitle={Article}, pdfauthor={Author}]{
    hyperref
} 

\begin{document}
    \title{A Full Minimal Coupling GW-BSE Framework for Circular Dichroism in Solids: Applications to Chiral 2D Perovskites}
    \author{Xian Xu}
    \affiliation{Department of Applied Physics, Yale University, New Haven, CT, 06511, USA}
    \author{Diana Y. Qiu}
    \email{diana.qiu@yale.edu} 
    \affiliation{Department of Materials Science, Yale University, New Haven, CT, 06511, USA}

    \date{\today}

    \begin{abstract}

    Circular dichroism (CD) and other chiroptical responses are a key probe of both chirality and momentum-space geometry in solids, but first-principles calculations are still challenging in periodic systems with strong exciton effects. Here, we develop a gauge-invariant first-principles framework for CD including exciton effects based on full minimal coupling (FMC) within the GW plus Bethe-Salpeter equation (GW-BSE) formalism. In contrast to standard multipole expansion and sum-over-states (SOS) approaches, which require careful gauge-fixing, converge slowly, and suffer origin ambiguities, FMC evaluates optical matrix elements directly at finite photon wavevector, naturally including intraband and near-degenerate transitions while placing electric‑dipole (ED), magnetic‑dipole (MD), and electric‑quadrupole (EQ) contributions on equal footing. Applied to two prototypical two-dimensional chiral hybrid perovskites, (S-NEA)\(_2\)PbBr\(_4\) and (S-MBA)\(_2\)PbI\(_4\), our calculations reveal that MD and EQ channels contribute equally to the CD signal.  Crucially, intraband and quasi‑degenerate transitions only captured within FMC can significantly modify CD spectra, especially in systems with dense band degeneracies.  The FMC framework, therefore, offers a computationally efficient and numerically robust way for predicting chiral optoelectronic phenomena in complex solids.

    \end{abstract}

    \maketitle

    Two-dimensional (2D) chiral hybrid organic-inorganic perovskites (HOIPs) have recently emerged as an exciting platform for chiroptoelectronics. In these materials, chiral molecular spacers are embedded into layered metal-halide perovskite structures\cite{Blum2024}, transferring chirality from the organic molecules to the inorganic sublattice \cite{Jana2020, Jana2021}. Structural chirality leads to an interplay of light polarization, charge-carrier spin, and crystal handedness, producing distinctive optical phenomena\cite{Long2020} such as circularly polarized light emission and detection \cite{Li2024}, chiral-induced spin selectivity (CISS) \cite{Lu2019, Lu2020}, spin-dependent photovoltaics\cite{Wang2021}, and chiral phonon-induced spin currents \cite{Kim2023}. Recent experiments report remarkably strong intrinsic circular dichroism (CD)---i.e. the differential absorption of left-circularly polarized (LCP) and right-circularly polarized (RCP) light, in chiral 2D lead-halide perovskites \cite{Lin2021, Dang2020, Ahn2020}, and \textit{ab initio} calculations show that this response is driven by an electron–hole exchange splitting of bright exciton states arising from the inorganic sublattice \cite{Li2024,Hautzinger2025, Sercel2025}. In addition, \textit{ab initio} calculations reveal that screening from organic cations contributes significantly to quasiparticle and optical spectra, reinforcing the role of many-body interactions in accurate calculations of these systems \cite{Filip2022,McArthur2023,Ziegler2022, Boeije2025, Leppert2024, Klymenko2024, Mihalyi2025}.
    
    However, \textit{ab initio} calculation of CD in such excitonic crystals, including 2D-HOIPs, poses significant theoretical challenges. Traditional approaches to optical activity were largely developed for molecules\cite{Rogers2024}, where light-matter coupling is treated either by direct evaluation of the position operator in time‑dependent density functional theory (DFT)\cite{Varsano2009, Yabana1999, Autschbach2011, Mattiat2021, Diedrich2003}, or by multipole expansions and sum‑over‑states (SOS) techniques \cite{Hidalgo2009, Warnke2012, Wang2023, Mahfouzi2025, Malashevich2010}.
    Directly applying these methods to extended periodic systems can lead to gauge ambiguities and convergence issues. For example, truncated SOS evaluations often yield results spuriously dependent on the choice of origin or on the gauge of the vector potential unless explicit gauge corrections are applied\cite{Ocana2023}, and naive velocity-gauge formulations of optical rotation exhibit apparent zero frequency divergences if intraband contributions are not treated properly\cite{Ocana2023}. Within the multipole expansion, computing orbital angular-momentum and electric-quadrupole matrix elements often requires either computationally intensive SOS calculations or numerically delicate wavefunction derivative techniques\cite{Malashevich2010}.
    These approaches have been implemented at the density functional theory (DFT) level for magnetic susceptibility\cite{Mauri1996} and nonlinear optics~\cite{SIPE1993, AVERSA1995, Souza2004}, and more recently, CD spectra of relatively simple chiral crystals such as CoSi, quartz, and Te, where excitons effects are less pronounced \cite{Multunas2023,Wang2023}. However, the mean field nature of conventional DFT fails to capture many-body effects, especially electron-hole correlations that are needed for accurate optical excitations\cite{Rohlfing2000, Onida2002}. Consequently, the CD spectra of complex chiral solid-state systems with strong exciton effects, such as 2D-HOIPs, remain incompletely understood.
    
    In this Letter, we introduce a Full Minimal Coupling (FMC) approach to CD in solids that combines the GW approximation for quasiparticle energies with the Bethe–Salpeter equation (BSE) for electron-hole interactions and apply it to understand the origin of CD in chiral HOIPs. The FMC method goes beyond the dipole approximation by explicitly incorporating the full light-matter interaction via minimal coupling, $\mathbf{p} \to \mathbf{p} + \frac{e}{c}\mathbf{A}$, in the excitonic Hamiltonian. By treating the perturbation to first order in the electromagnetic wavevector $\mathbf{q}$ (i.e. including spatial-dispersion effects), this approach inherently captures terms equivalent to the electric dipole (ED), magnetic dipole (MD), and electric quadrupole (EQ) contributions on an equal footing. Importantly, the FMC formulation is manifestly gauge-invariant and free of the divergences that plague naive SOS and multipole-expansion treatments, since it requires neither an arbitrary origin choice nor incomplete summation of states. We benchmark the FMC approach against SOS in the multi-pole expansion for two 2D‑HOIPs, (S‑NEA)$_{2}$PbBr$_{4}$ (abbreviated S‑NPB; NEA = 1‑(1‑naphthyl)ethylammonium) and (S‑MBA)$_{2}$PbI$_{4}$ (abbreviated S‑MPI; MBA = $\alpha$-methylbenzylamine), and show improved numerical stability and faster convergence for materials with large numbers of nearly degenerate bands. 



    \textit{Ab initio} calculations of optical response in crystals generally start by treating the vector potential of the incident light $\mathbf{A}=\mathbf{A}_{0}e^{i \mathbf{q} \cdot \mathbf{r}}$ as a perturbation, with the minimal coupling interaction Hamiltonian $H_{int}=-\frac{e}{m_{e}}\mathbf{p}\cdot \mathbf{A}$, where $\mathbf{p}$ is the electron momentum operator, $\mathbf{r}$ the position operator and $\mathbf{q}$ the photon wave vector. In the long wavelength limit, typical for linear response, $\mathbf{q}\!\to\!0$ so that $\mathbf{A}$ is approximated as spatially uniform and only the ED term is retained. CD, however, vanishes in this limit, as a finite $\mathbf{q}$ is required to retain first-order spatial-dispersion effects that are odd under parity.
    
    We first summarize the conventional SOS approach\cite{Deilmann2020, KOHN1959, Wang2023, Multunas2023, Hidalgo2009}. Expanding the interaction Hamiltonian\cite{Malashevich2010} to the first order in $\mathbf{q}$ gives
    \begin{equation}
        H_{\text{int}}=-\frac{e}{m_{e}}\,\mathbf{p}\!\cdot\!\mathbf{A}_{0}-\frac{e}{2m_{e}}
        \,\mathbf{L}\!\cdot\!(\nabla\!\times\!\mathbf{A}) -\frac{i e }{2m_{e}}\mathbf{A}\!\cdot\!\mathbf{Q}\!\cdot\!\mathbf{q},
    \end{equation}
    where $\mathbf{L}=\mathbf{r}\times\mathbf{p}$ is the orbital angular momentum and $\mathbf{Q}=\mathbf{r}\otimes\mathbf{p}+\mathbf{p}\otimes\mathbf{r}$ is the electric quadrupole. We neglected the spin angular momentum contribution here for simplicity, as in the systems studied, its magnitude is less than 2\% of the angular momentum \cite{Tsirkin2018, Malashevich2010} (also see SI). Within Fermi's golden rule, CD arises from cross terms between ED and EQ or ED and MD interaction \cite{PEDERSEN1995}. For molecules in the gas phase, only the ED-MD cross term survives after orientational averaging\cite{Govorov2010}, but in periodic solids, the EQ term remains to the same order as the MD\cite{Multunas2023}. 

    
    Evaluating $\mathbf{L}$ and $\mathbf{Q}$ in a Bloch basis is nontrivial because they involve $\mathbf{r}$. A common approximation converts position matrix elements, between Bloch states ($|n\mathbf{k}\rangle$,$|m\mathbf{k}\rangle$), to momentum matrix elements using\cite{Fei2020}
    \begin{equation}
        \langle n \mathbf{k}|[\mathbf{r}, H]|m \mathbf{k}\rangle=\langle n \mathbf{k}| r|m \mathbf{k}\rangle\left[E_{m k}-E_{n k}\right]=\frac{i \hbar}{m_e}\langle n \mathbf{k}| \mathbf{p}|m \mathbf{k}\rangle
    \end{equation}
   Components of $\mathbf{L}$ and $\mathbf{Q}$ are then computed via a SOS expression\cite{Deilmann2020},
    \begin{equation}
        \langle v\mathbf{k}\lvert p_{x}r_{z}\rvert c\mathbf{k}\rangle = -\frac{i\hbar}{m_{e}}
        \sum_{l}\frac{ \langle v\mathbf{k}\lvert p_{x}\rvert l\mathbf{k}\rangle \langle
        l\mathbf{k}\lvert p_{z}\rvert c\mathbf{k}\rangle }{ E_{l\mathbf{k}}- E_{c\mathbf{k}}}
        .
    \end{equation}
    The required momentum matrix elements can be obtained (i) exactly from the commutator $\mathbf{p}=\frac{m_e}{i\hbar}[H,\mathbf{r}]$, introducing a finite difference between wavefunctions at nearby \textbf{k}-points, or (ii) by approximating $\mathbf{p}=-i\hbar\nabla$, which neglects non-local contributions to the Hamiltonian\cite{Marini2001,sorhab2001}.
    
To include excitons, we solve the BSE in the Tamm-Dancoff approximation, where the eigensolution corresponding to an exciton state takes the form 
$|S\rangle = \sum_{vc\mathbf{k}}A^S_{vc\mathbf{k}}|vc\mathbf{k}\rangle$, with the envelope $A_{v c \boldsymbol{k}}^S$ weighting the electron state $c\mathbf{k}$ and the hole state $v\mathbf{k}$. In the excitonic basis, the ED, MD and EQ matrix elements can be written as $\langle0| p_{x}|S\rangle=\sum_{c v k}A_{v c k}^{S}\langle v k| p_{x}|c k\rangle$, $\langle0| L_{x}|S\rangle=\sum_{c v k}A_{v c k}^{S}\langle v k| L_{x}|c k\rangle$ and $\langle 0| Q_{yz}|S\rangle=\sum_{c v k}A_{v c k}^{S}\langle v k| Q_{yz}|c k\rangle$.
    Using Fermi's golden rule, the imaginary part of the dielectric function, $\epsilon'' (\omega)$, which is proportion to absorption, is  
    \begin{equation}
        \varepsilon^{\prime \prime}(\omega)=\frac{8 \pi^2 e^2}{\omega^2} \sum_{v c k}\left|X_{ED}^{v c k}+X_{MD}^{v c k} + X_{EQ}^{v c k} \right|^2 \delta\left(\omega-\omega_{v c k}\right)
    \end{equation}  
    in the independent‑particle approximation, and
    \begin{equation}
        \varepsilon^{\prime \prime}(\omega)=\frac{8 \pi^2 e^2}{\omega^2} \sum_{S}\left|X_{ED}^{S}+X_{MD}^{S} + X_{EQ}^{S} \right|^2 \delta\left(\omega-\Omega_{S}\right)
    \end{equation}
    with excitonic effects included, where
    $X_{ED}^{v c \boldsymbol{k}}=\hat{A} \cdot\langle v \boldsymbol{k}| {\boldsymbol{p}}|c \boldsymbol{k}\rangle$, $X_{MD}^{v c \boldsymbol{k}}=\frac{1}{2}\langle v \boldsymbol{k}| {\boldsymbol{L}} \cdot(\nabla \times \hat{A})|c \boldsymbol{k}\rangle$ and $X_{EQ}^{v c \boldsymbol{k}}=\frac{1}{2}\langle v \boldsymbol{k}| i\hat{A}\!\cdot\!\mathbf{Q}\!\cdot\!\mathbf{q}|c \boldsymbol{k}\rangle$. Here, $\hat{A}=\boldsymbol{A} /|\boldsymbol{A}|$ is the direction of the vector potential, and $\omega_{v c k}$ and $\Omega_{S}$ are the excitation energies of a free electron-hole pair and and exciton, respectively. The excitonic quantities $X_{ED}^{S}$, $X_{MD}^{S}$, and $X_{EQ}^{S}$ are defined analogously.

    \begin{figure}
        \centering
        \includegraphics[width=\linewidth]{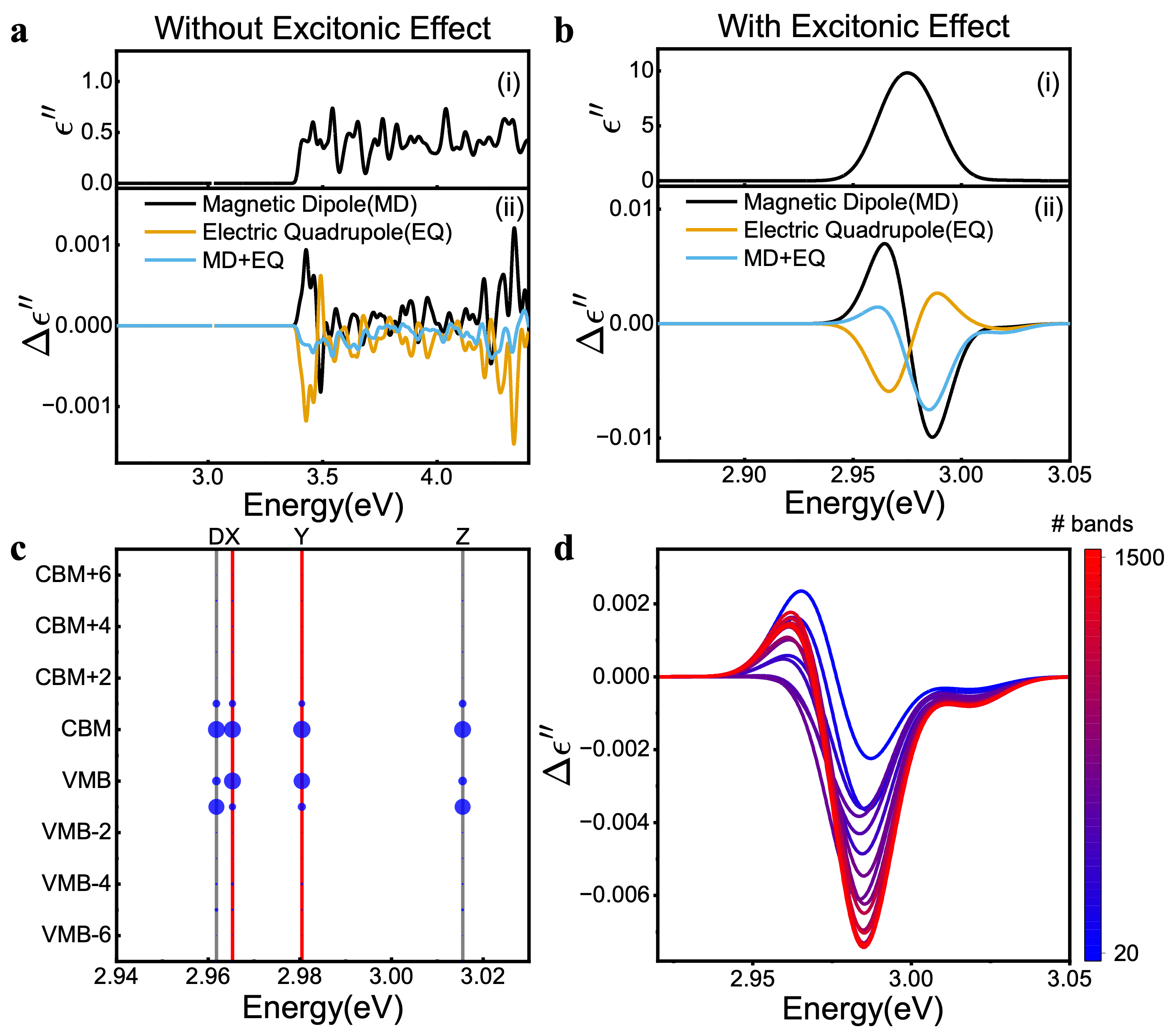}
        \caption{\textbf{SOS CD calculations for S-NPB.}
        \textbf{a} Spectra at the GW independent-particle level. showing (i) the imaginary part of the dielectric function \(\epsilon''\), and (ii) the differential imaginary response computed with  MD-only, EQ-only, and both contributions. \textbf{b} Same as \textbf{a} at the GW-BSE level including excitons. \textbf{c}, Contribution of independent-particle bands (VBM=highest valence band; CBM=lowest conduction band) to the four lowest energy excitons (D, X, Y, Z); the optically bright X and Y states are highlighted in red. The size of the circles scales with $\sum_{c\mathbf{k}}|A^S_{vc\mathbf{k}}|^2$ for valence states or $\sum_{v\mathbf{k}}|A^S_{vc\mathbf{k}}|^2$ for conduction states. \textbf{d}, Convergence of the excitonic CD spectrum with respect to the number of bands in the SOS summation. }
        \label{fig1}
    \end{figure}

    While formally exact, the SOS approach described above suffers from slow convergence with the number of states \cite{Multunas2023}, divergent energy denominators for intraband transitions and interband transitions between nearly degenerate bands, and gauge/origin ambiguities, since truncated sums spuriously depend on the choice of unit‑cell origin or vector-potential gauge unless explicit gauge corrections are applied\cite{Ocana2023}. These challenges severely limit the practical accuracy and reliability of SOS-based CD calculations in complex materials with dense band manifolds. To retain intra‑band terms, one may adopt Blount’s formulation \cite{BLOUNT1962}, which connects the position operator in solids with the Berry connection\cite{Souza2004}. This length-gauge method has been widely used for nonlinear optical responses in solids\cite{AVERSA1995, SIPE1993, Ruan2024}.

    However, while Blount’s formulation removes SOS divergences and includes intraband and degenerate-band transition contributions, it introduces new challenges. It depends on numerical derivatives $\nabla_{\mathbf{k}}\ket{u_{n\mathbf{k}}}$, requiring (i) a dense $\mathbf{k}$‑mesh and (ii) a smooth gauge to remove the arbitrary random phases from diagonalizing the Hamiltonian. Gauge-smoothing techniques, such as parallel transport or Wannier gauges\cite{Marzari1997,Souza2004}, can regularize the single-particle phases, but become increasingly numerically challenging for excitons because the exciton envelop $A_{vc\mathbf{k}}^S$ carries additional \textbf{k}- and S-dependent phases, introducing extra covariant derivatives into the formalism~\cite{Souza2004}. Moreover, the dense k-mesh required to converge finite-element derivatives can be computationally prohibitive for BSE calculations that scale as $O(N^5)$ with the number of states.

    To overcome these difficulties, we develop the Full Minimal‑Coupling (FMC) method, which bypasses both the SOS summation and any explicit derivatives. The FMC method starts from the minimal‑coupling Hamiltonian without expanding it in the wavevector $\mathbf{q}$ so that the matrix element within first order perturbation theory is $\bra{v\mathbf{k}+\mathbf{q}}\! e^{i\mathbf{q}\cdot\mathbf{r}}\mathbf{p}_{x}\ket{c\mathbf{k}}$. Thus, the FMC formalism retains the position operator explicitly, capturing the requisite derivative information through the small but finite momentum transfer $\mathbf{q}$ from the photon. When electron–hole interactions are included, the corresponding excitonic matrix element becomes
    \begin{equation}
        \bra{0}e^{i\mathbf{q}\cdot\mathbf{r}}\mathbf{p}_{x}\ket{S}=\sum_{vck}A^{S}
        _{vc\mathbf{kq}}\bra{v\mathbf{k}+\mathbf{q}} e^{i\mathbf{q}\cdot\mathbf{r}}\mathbf{p}_{x}\ket{c\mathbf{k}}
    \end{equation}
    where $A^{S}_{vck\mathbf{q}}$ is the exciton wavefunction envelope of a state with a small finite momentum $\mathbf{q}$ (see SI for planewave expansion). The dielectric function can be constructed as usual through Fermi's golden rule.
    The only approximation that remains is the replacement of the canonical momentum by $-i\hbar\nabla$ to avoid numerical complications arising from applying two separate momentum shifts. Consequently, nonlocal contributions to $\mathbf{p}$ from the pseudopotential are omitted\cite{Marzari1997,sorhab2001}. We explore the effect of neglecting this term in the SI.


    We benchmark our approach on two prototypical chiral 2D-HOIPs S-NPB \cite{Hautzinger2025} and S-MPI\cite{Ahn2020}. Ground‑state wavefunctions are obtained from DFT calculations within the generalized gradient approximation (GGA) \cite{Perdew1996} as implemented in Quantum Espresso \cite{Giannozzi2009} within a fully-relativistic spinor formalism using norm‑conserving pseudopotentials \cite{Hamann2013}. Quasiparticle energies are corrected at the one‑shot $G_{0}W_{0}$ level with BerkeleyGW \cite{Deslippe2012, HYBERTSEN1986}. Excitonic effects are included by solving the BSE\cite{Deslippe2012, Rohlfing2000}. CD spectra are then evaluated via SOS and FMC at the independent‑particle level and with electron–hole interactions. Detailed calculation parameters are reported in the SI.

    We start with the chiroptical response within SOS. Fig.~\ref{fig1} shows the imaginary part of the dielectric function ($\epsilon''$) and differential dielectric function ($\Delta\epsilon'')$ under LCP and RCP excitation with a uniform empirical broadening of 10 meV. We find that the MD and EQ components have comparable magnitudes but opposite signs, confirming that both are essential for a quantitative CD description in solids. When combined, their partial cancellation produces a substantially smaller total CD signal; this behavior persists with excitonic effects (Fig.~\ref{fig1}b). After including exciton affects, the low-energy spectrum of S-NPB is dominated by two bright excitons states, as discussed in previous work \cite{Li2024, Sercel2025}. Previous CD analysis only considered the MD contribution, producing a symmetric derivative-like lineshape(Cotton effect). Since the MD and EQ contributions have opposite signs and the lowest-energy bright exciton states are split by only 17 meV (Fig.~\ref{fig1}c), a realistic spectral broadening of 10 meV partially merges these features causing the Cotton effect to become less prominent and giving rise to an asymmetric peak structure whose lineshape converges slowly and nonmonotonically with the number of bands included in the SOS (Fig.~\ref{fig1}d). While earlier work from us and others \cite{Li2024, Sercel2025} reported a symmetric lineshape, informed by theoretical calculations that neglected the EQ term, closer analysis of some published experimental spectra do reveal an asymmetric lineshape that may have been overlooked\cite{Sercel2025, Hautzinger2025}.

    \begin{figure}
        \centering
        \includegraphics[width=\linewidth]{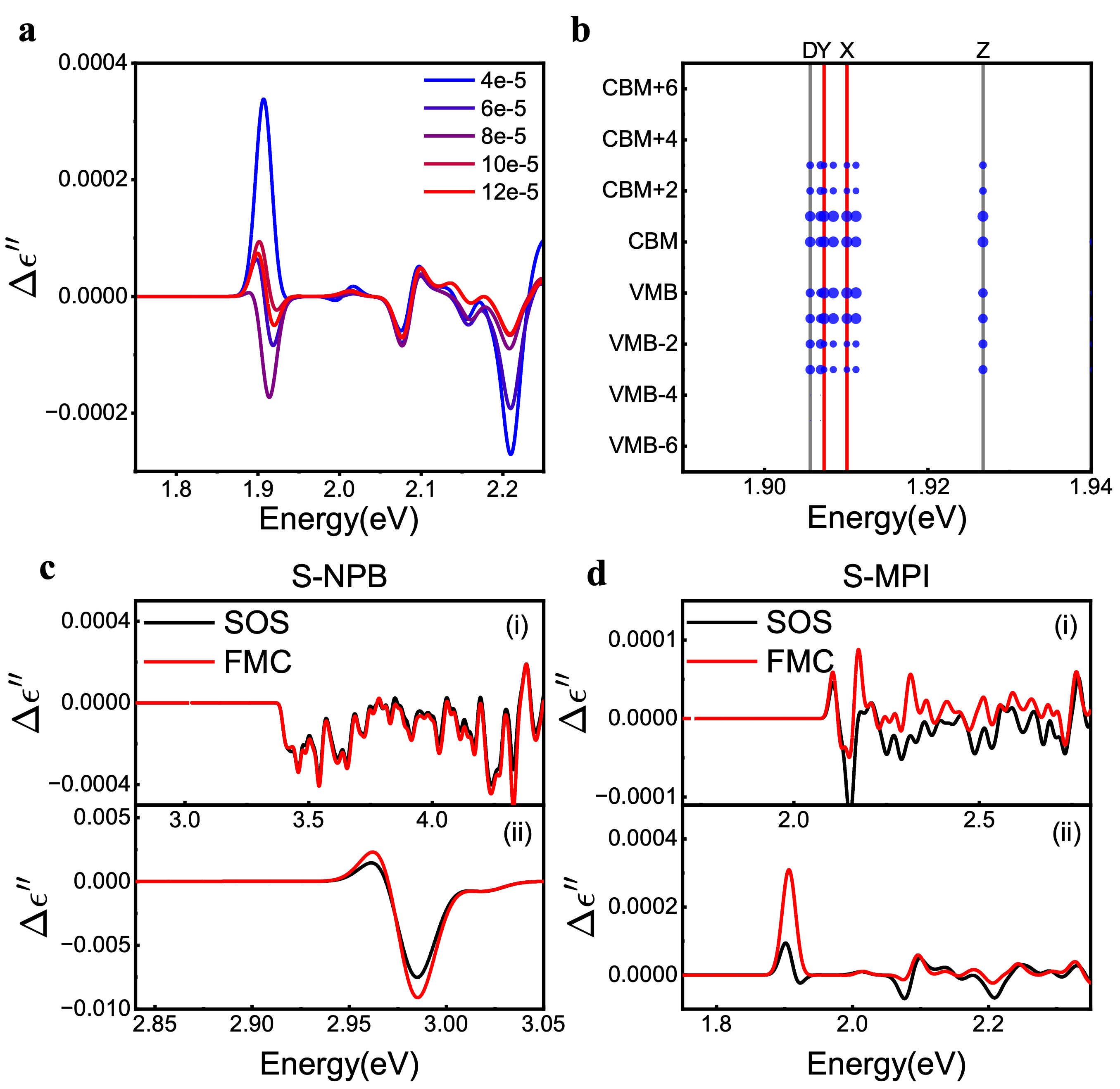}
        \caption{\textbf{FMC CD calculations for 2D-HOIPs.} \textbf{a}, CD spectra for S‑MPI (with excitons) obtained via SOS using different energy thresholds (Rydbergs) to identify degenerate states. \textbf{b}, Contributions of independent‑particle bands to excitons near the first bright peak. X, Y, Z label the brightest states for the corresponding linear polarizations; D denotes the lowest dark state. Bright X and Y are highlighted in red. \textbf{c,d}, CD spectra from SOS and FMC at the independent‑particle level (lower panels) and including excitonic effects (upper panels) for \textbf{c} S-NPB and \textbf{c} S-MPI.}
        \label{fig2}
    \end{figure}

    To understand the fine structure and asymmetric line shape in S‑NPB, we decompose the excitonic character in Fig.~\ref{fig1}(c). Consistent with previous work, there are two low-energy in-plane excitons that are bright under excitation by in-plane polarized light (X, Y), as well as lower energy dark state (D) and a higher energy out‑of‑plane state (Z) \cite{Sercel2025}. These excitons are primarily composed of the two highest valence bands and two lowest conduction bands, which have the character of the inorganic octahedra (see SI). Thus, molecular cation states do not contribute to the exciton wave function or to linear absorption, which is dominated by the ED term. However, in the higher‑order MD and EQ channels, the SOS brings molecular states into the chiroptical response. Indeed, the convergence of the CD spectrum within the SOS framework (Fig.~\ref{fig1}(d)) shows that retaining only the inorganic bands results in a symmetric CD spectrum similar to a theoretical four band model\cite{Sercel2025}, while progressively adding molecular states introduces asymmetry that converges nonmonotonically. More than 1,000 bands are required to converge the CD spectrum.

    \begin{figure}
        \centering
        \includegraphics[width=\linewidth]{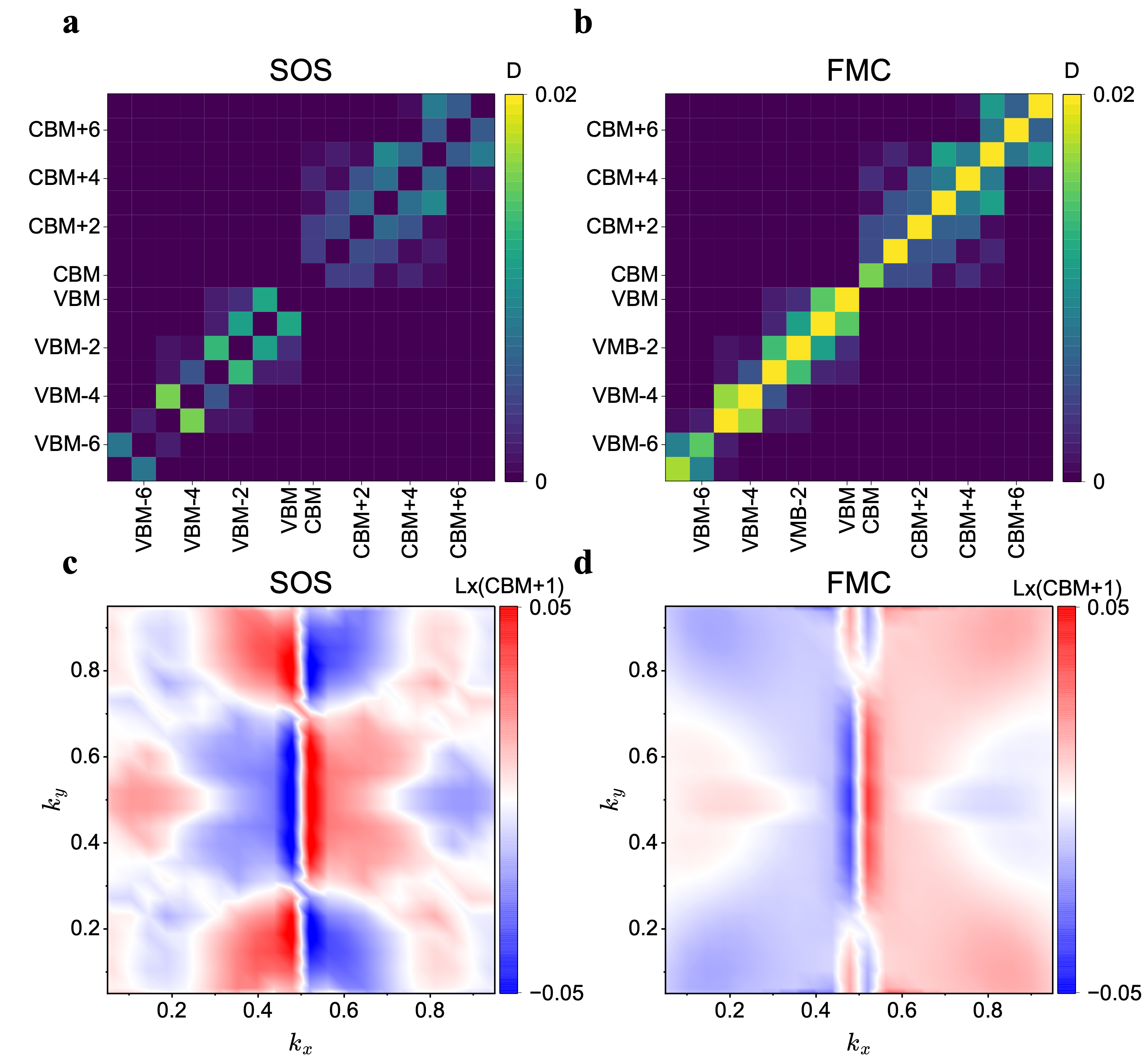}
        \caption{\textbf{Analysis of SOS and FMC.} \textbf{a,b}. Spatially dependent part of the transition matrix element D of S-NPB from the SOS and FMC approaches.  \textbf{c,d}. Diagonal elements of the electron orbital-angular-momentum matrix for S-MPI evaluated  with (\textbf{c}) SOS and (\textbf{d}) FMC.}
        \label{fig3}
    \end{figure}
    
    We now turn to a different chiral HOIP, S-MPI. While convergence of SOS is merely slow for S-NPB, a naive SOS implementation breaks down completely for S-MPI. S-MPI is a chiral double‑layer perovskite whose unit cell contains two layers of PbI$_{4}$ octahedra separated by two layers of molecular spacers. The octahedral layers interact only weakly, resulting in a band structure that is nearly (but not exactly) degenerate at every point (See SI for band structrue). Transitions among these nearly degenerate states contribute strongly to the CD response, but evaluation of the energy denominator in the SOS approach (Eq. 3) becomes numerically unstable. Fig.~\ref{fig2}(a) shows a naive SOS approach where we neglect contributions from degenerate states: the CD spectrum oscillates dramatically with the energy threshold used to define degenerate states. Fig.~\ref{fig2}(b) shows the spectrum of bright and dark excitons states and their composition. We see a manifold of dark states and linear polarized bright states, very similar to S-MPI, but with a smaller energy splitting due to a smaller exchange interaction.
    
     To avoid these numerical challenges in the SOS approach, we turn to the FMC approach. Fig.~\ref{fig2}(c,d) compare SOS and FMC CD spectra for S‑NPB and S‑MPI, with and without excitons. For S-NPB, FMC slightly enhances the magnitude of the excitonic and non-interacting CD spectra, but qualitative features remain the same. For S-MPI, on the-other-hand, FMC produces a qualitatively reliable spectrum, whereas SOS results are numerically unstable and can fluctuate dramatically with computational parameters (the plotted SOS curves are illustrative only).

    Beyond stability and efficiency, FMC naturally includes intraband contributions neglected by SOS. Fig.~\ref{fig3}(a,b) compare the spatially-dependent part of the transition matrix element $D = \sum_{k}|\bra{mk}e^{iqr}-1\ket{nk}|^2$ for S-NPB. SOS neglects diagonal (intraband) elements, while FMC explicitly captures sizable diagonal terms. This intraband contribution has the form of the Berry connection, highlighting the role of quantum geometry in CD, which is included naturally through the FMC approach.

   Finally, we note that FMC is robust not only for CD (interband transitions) but also for band‑resolved quantities such as orbital angular momentum. In Fig.~\ref{fig3}(c,d), the diagonal elements of the orbital‑angular‑momentum matrix $L_{nn}$ for S-MPI differ markedly between SOS and FMC. Similarly to the CD, the SOS exhibits unstable fluctuations that change depending on numerical parameters, whereas the FMC data are smooth and physically consistent.

    In summary, we developed a gauge‑invariant FMC framework for CD in chiral solids that integrates GW quasiparticles with BSE excitons. Compared with conventional multipole SOS approach, FMC eliminates slow convergence with respect to empty states and divergences from band degneracies, while explicitly including physics from intraband transitions. Benchmarks on S-NPB and S-MPI reveal that MD and EQ channels are both essential. FMC further shows that intraband and quasi‑degenerate transitions can dramatically reshape the CD spectra, highlighting the role of $\mathbf{k}$-derivative (Berry connection-like) terms in chiroptical physics. These findings establish FMC as a robust and reliable framework for studying CD in complex chiral materials and provide a foundation for future investigations into chiral optoelectronics.

    \text { Acknowledgments}: This work was primarily supported by the National Science Foundation Division of Chemistry under award number CHE-2412412. Development of the BerkeleyGW code was supported by the Center for Computational Study of Excited-State Phenomena in Energy Materials (C2SEPEM) at the Lawrence Berkeley National Laboratory, funded by the U.S. Department of Energy, Office of Science, Basic Energy Sciences, Materials Sciences and Engineering Division, under Contract DE-AC02-05CH11231 (D.Y.Q.). The calculations used resources of the National Energy Research Scientific Computing (NERSC), a DOE Office of Science User Facility operated under contract no. DE-AC02-05CH11231, under award BES-ERCAP-0031507 and BES-ERCAP-0027380; and the Texas Advanced Computing Center (TACC) at The University of Texas at Austin. D.Y.Q. would like to thank M. Filip and S. Refaely-Abramson for helpful discussions.

    \bibliography{main}
\end{document}